\documentclass[a4paper,12pt]{article}
\usepackage{mathrsfs}
\usepackage{graphicx}
\usepackage{epstopdf}
\usepackage{subfigure}
\usepackage{color}
\usepackage{ulem}
\usepackage{float}
\usepackage{amsmath} 
\begin{document}

\hoffset = -1truecm \voffset = -2truecm \baselineskip = 10 mm
\title{\bf A possible evidence of pion condensation}

\author{Wei Zhu$^{a,b}$\footnote{Corresponding author, E-mail: wzhu@phy.ecnu.edu.cn}, Yu-Chen Tang$^c$, Lei Feng$^c$ and Feng-yao Hou$^d$
\\\\
         \normalsize $^a$Center for Fundamental Physics and School of Mathematics and Physics,\\
Hubei Polytechnic University, Huangshi 435003, P.R. China\\
         \normalsize $^b$Department of Physics, East China Normal University, Shanghai 200241, P.R. China\\
         \normalsize $^c$Key Laboratory of Dark Matter and Space Astronomy, Purple Mountain Observatory,\\
         \normalsize Chinese Academy of Sciences, Nanjing 210008, P.R. China\\
         \normalsize $^d$Institute of High Energy Physics, Chinese Academy of Sciences, Beijing, P.R. China,\\
Spallation Neutron Source Science Center, Dongguan, P.R. China\\
         }
\date{}

\newpage
\maketitle

\vskip 3truecm

\begin{abstract}

This work demonstrates that once a large number of pion is condensed in a high-energy hadron collision, the gamma-ray spectrum from $\pi^0$ decay takes on a typical broken power-law shape, which has been documented in many astronomical observations, but we have not yet recognized it. We show that this pion condensation is caused by a large number of soft gluons condensed in protons.

\end{abstract}

{\bf keywords}: Gamma-ray astronomy; Nuclear physics; Ultra-high-energy cosmic radiation; Pion condensation

\newpage

\vskip 1truecm

\section{INTRODUCTION}

    Pion condensation is a special state of matter, where pions condense in the same quantum state
under certain extreme conditions. This concept was first introduced by physicist A. B. Migdal, who discussed the possibility of this phenomenon in nuclear matter in his 1971 paper $^{[1]}$. With the advancement of experimental techniques and theoretical models, the study of pion condensation has made remarkable progress. Modern studies have focused on the possibility and characterization of pion condensation under extreme conditions (e.g., in neutron star and heavy ion collisions). It has now evolved into a multidisciplinary cross-study area covering high-density QCD theory, numerical simulations, heavy-ion collision experiments and astrophysical observations.
The confirmation of pion condensation will not only be an important validation of fundamental physical theories, but also have far-reaching implications for several branches of physics, including particle physics, astrophysics, condensed matter physics and nuclear physics. Unfortunately, there is currently no solid experimental evidence confirming the existence of pion condensation.

   In this paper, we demonstrate that once a large number of pions
condense in high-energy hadron collisions, the gamma-ray spectra produced by $\pi^0$-decay will take a typical broken power law (BPL) shape, which has already existed in many astronomical observations, but we have not yet recognized it.

    High-energy $pp$ collisions produce a large number of secondary hadrons in the central region of rapidity, which are
dominated by pions. As observed in high-energy hadron collisions, the particle yield in the central region increases slowly with $\ln \sqrt{S}$, and more of the collision energy $\sqrt{S}$ is converted into the kinetic energy of the thermal motion of the new particles in the center-of-mass (C.M.) system. This scenario favors the formation of quark-gluon plasma (QGP), but it does not meet the high-density and low-temperature conditions required for pion condensation.

     Pion condensation is defined as the pion field obtaining a non-zero
vacuum expectation value, because a large number of pions accumulate in the ground state.High density and extremely low temperature are two conditions that cause pion condensation.
Obviously, in order to condense pions, the temperature in the collision center region must be effectively reduced. We find that this method hides the QCD structure of the proton.

    A series of QCD evolution equations studies have found that at high energies, the gluon distribution in the nucleon exhibits
chaotic behaviour, which induces strong shadowing and antishadowing effects, the latter leading to the aggregation of innumerable gluons in a narrow phase space defined by the critical momentum $(x_c, k_c)$. This is gluon condensation (Figure 1) $^{[2,3,4,5]}$.

\begin{figure}[H]
  \begin{center}
   \includegraphics[width=0.6\textwidth]{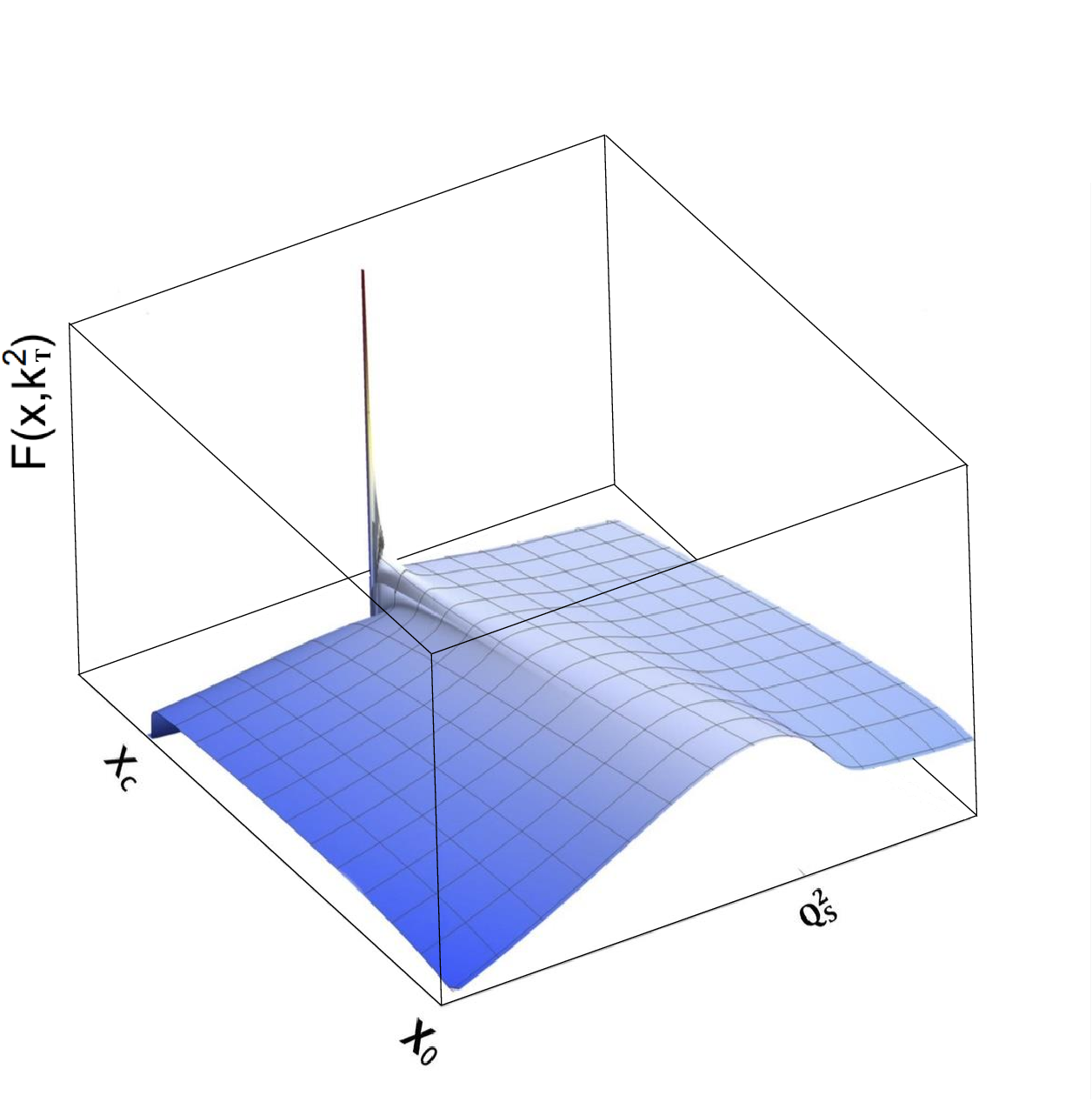}
   \parbox{15.5cm}{\small{\bf Fig.1.}  A schematic solution of a QCD evolution equation $^{[3]}$, which shows the evolution of
        transverse momentum dependent distribution $F(x,k^2_T)$ of gluons in proton from color glass condensation
        (CGC) $^{[6]}$ at $x_0$ to gluon condensation at $x_c$. Note that all gluons with $x<x_c$ are stacked at $(x_c,k_c)$.
}\label{fig:1}
  \end{center}
\end{figure}

    Although the Higgs boson is responsible for producing the current quark mass, recent studies have shown that the
hadron mass is mainly derived from the gluon field $^{[7]}$. Therefore, gluon interactions dominate the multi-production processes in high-energy proton collisions. Once the collision energy exceeds the threshold value corresponding to $x_c$, a large number of condensed gluons rapidly accumulate in the central region. In this case the pion yield will dramatically increase.
Energy conservation requests

\begin{equation}
    \begin{split}
        \frac{1}{2}\sqrt{S} &= N_{\pi,a}m_{\pi}+AT_a ~~\text{without gluon condensation}\\
        &= N_{\pi,b}m_{\pi}+AT_b, ~~\text{with gluon condensation}
    \end{split}
    \label{eq1}
\end{equation}

where $A$ is a constant, $T$ is the absolute temperature in Kelvin and the other half of the C.M. energy $\sqrt{S}$ was taken by the leading particles $^{[11]}$. One can find that if
$N_{\pi,b}\gg N_{\pi,a}$, we have  $T_b\ll T_b$. Therefore, under the same collision energy, the thermal motion energy of pions must be suppressed, resulting in cooling. It seems that an infinite number of condensed gluons could create an infinite number of pion, however, conservation of energy allows only a maximum number of pion to be created since pion has mass, i.e., almost all kinetic energy is used to produce static particles. This results in the newly formed pions having almost no relative momentum and $T_b\rightarrow 0$, which produces a maximum value of $N_{\pi}$. Consequently, a dense and low-temperature physical environment conducive to the formation of pion condensation is established in the central region of rapidity.

    The next question is what range the temperature in the collision center region should drop to for pion condensation to occur?
Okorokov found BEC correlations between particles with low relative momentum in high-energy hadronic
collisions, although it is not full boson condensation since only a small fraction of low relative momentum pions are correlated $^{[8,9,10]}$.This discovery implies that bosons with relatively low thermal motions at high densities can also produce boson correlations at temperatures no lower than the critical BEC temperature. So we assume that the relative momentum of the pion in the condensation state is much smaller than the pion  mass. Thus, the condensed pions can be approximated as stationary
in the C.M. system, while in Lab. they have the same Lorentz factor $\gamma$ (Figure 2).

\begin{figure}[H]
  \begin{center}
   \includegraphics[width=0.5\textwidth]{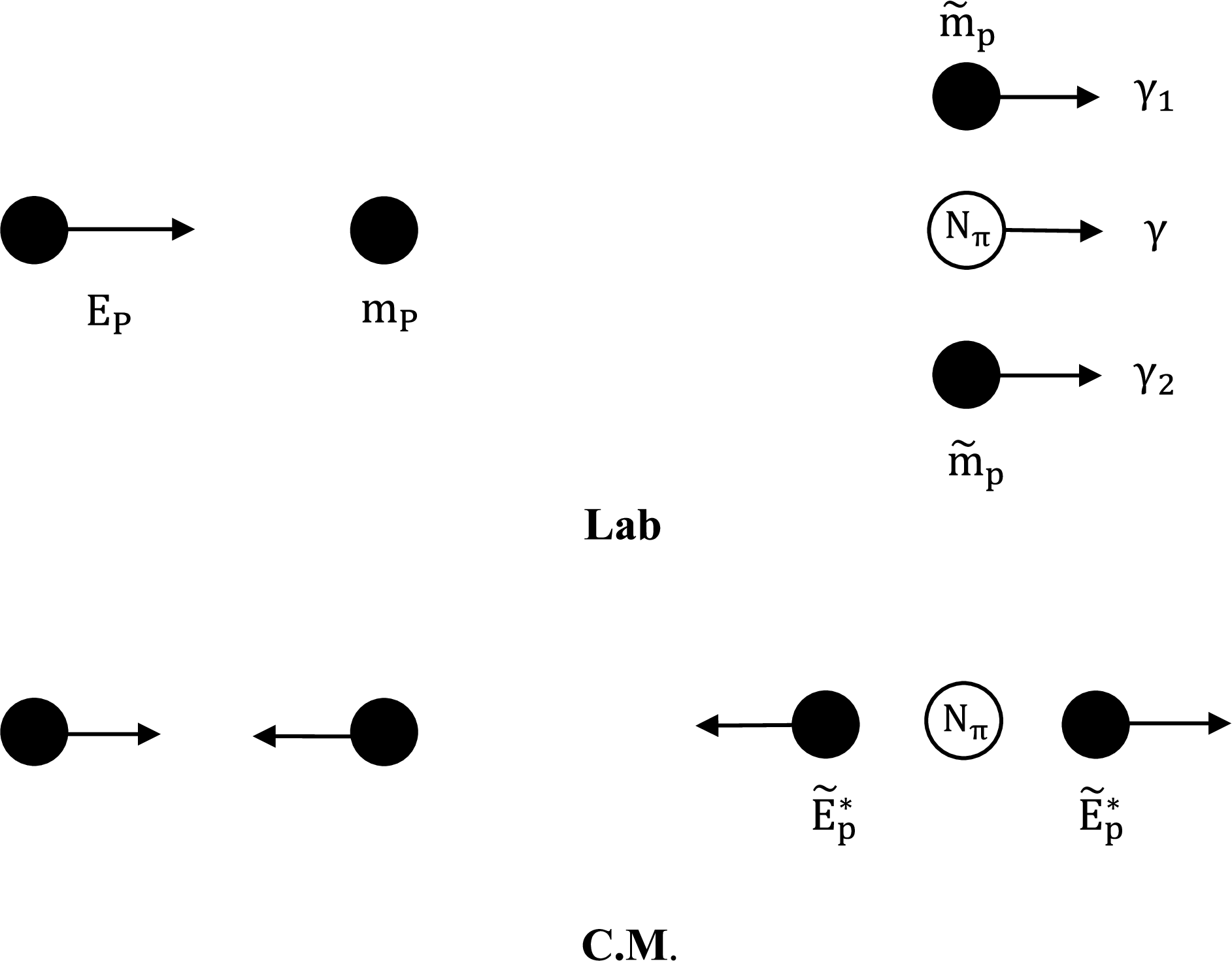}
   \parbox{15.5cm}{\small{\bf Fig.2.} $pp\rightarrow N_{\pi}$ in the different systems, where the pions are condensed.
}\label{fig:2}
  \end{center}
\end{figure}

    We simplify the process $pp\rightarrow X$ to $pp\rightarrow pp+N_\pi$. According to Figure 2 we write energy conservation

\begin{equation}
    E_p+m_p=\tilde{m}_p\gamma_1+\tilde{m}_p\gamma_2+N_{\pi}m_{\pi}\gamma,  \label{eq2}
\end{equation}
where $\tilde{m}_p$ marks the leading particle and
$\gamma_i$ is the Lorentz factor. Note that the leading particle carries about half of the total energy. On the other hand,
the square of relativistic invariant total energy $S=(p_1+p_2)^2$ in the two systems is
\begin{equation}
    S=2m_p^2+2E_pm_p=(2\tilde{E}^*_p+N_{\pi}m_{\pi})^2. \label{eq3}
\end{equation}

    The proportion of the total collision energy occupied by the leading particles and the central particles is independent
of the selection of the reference frame. Using the following empirical relation $^{[11]}$, we remove the physical quantities about the leading particles in Equations \ref{eq2} and \ref{eq3}
\begin{equation}
    2\tilde{E}^*_p=\left(\frac{1}{k}-1\right)N_{\pi}m_{\pi},
    \label{eq4}
\end{equation}
and
\begin{equation}
    \tilde{m}_p\gamma_1+\tilde{m}_p\gamma_2=\left(\frac{1}{k}-1\right)N_{\pi}m_{\pi}\gamma, \label{eq5}
\end{equation}
$k\simeq 1/2$ is the inelasticity.
Thus, we have the relations among $N_\pi$, $E_{\pi}$ and $N_p$
\begin{equation}
    \ln N_{\pi}=0.5\ln (E_p/{\rm GeV})+a, ~~\ln N_{\pi}=\ln (E_{\pi}/{\rm GeV})+b,  \label{eq6}
\end{equation}
$$~~ with~E_{\pi}
\in [E_{\pi}^{GC},E_{\pi}^{max}], $$where $a\equiv 0.5\ln (2m_p/{\rm GeV})-\ln (m_{\pi}/{\rm GeV})+\ln k$
and $b\equiv \ln (2m_p/{\rm GeV})-2\ln (m_{\pi}/{\rm GeV})+\ln k$. They are straight lines in double logarithmic coordinates, called
the power law (PL) (Figure 3a). This is the result of the condensation of a large number of pions in the $pp$ collision (see Figure 2), therefore,
we regard it as a replantation of pion condensation.

\begin{figure}[H]
	\begin{center}
		\includegraphics[width=0.8\textwidth]{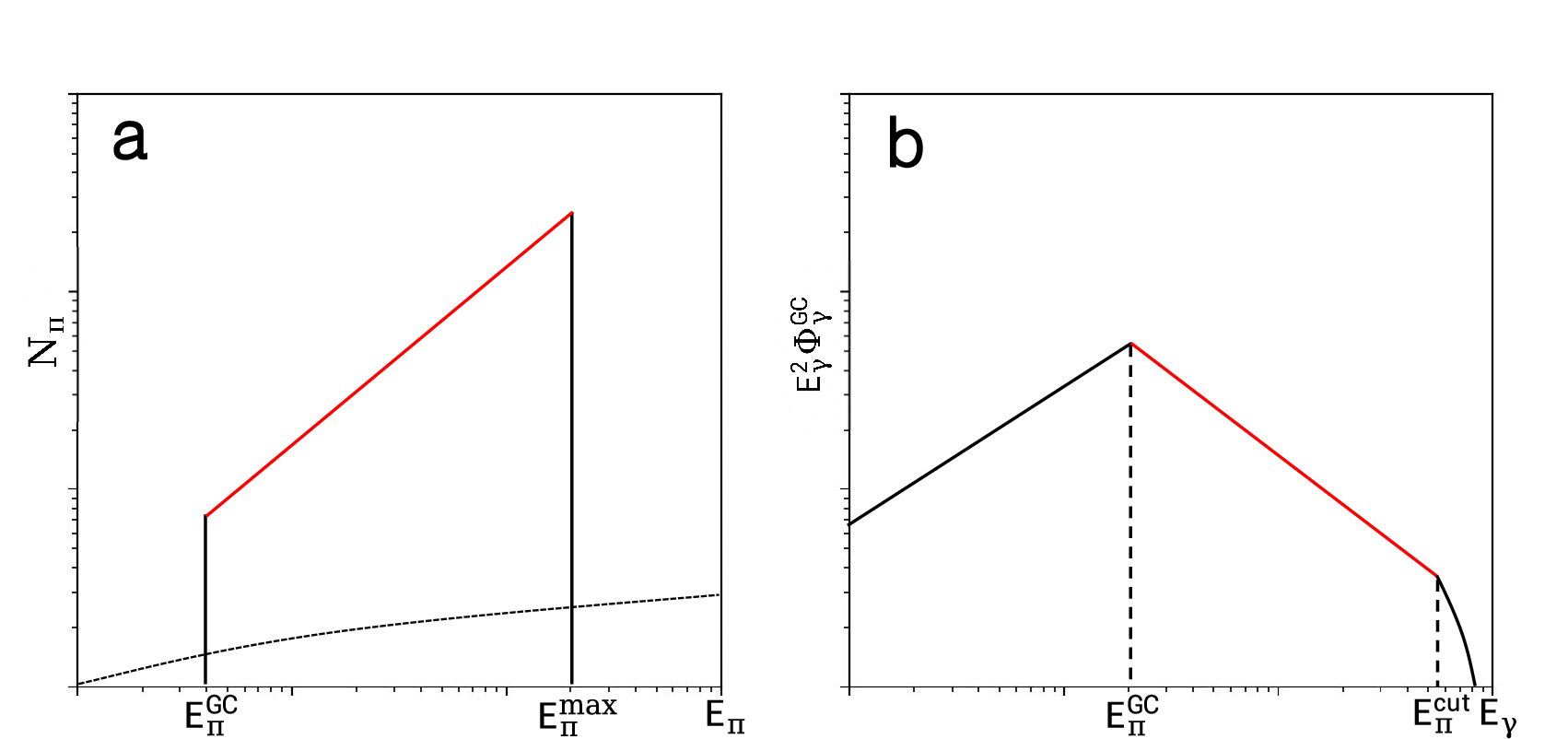} 
  
		\parbox{15.5cm}{\small{\bf Fig.3.}(a) The solid curve represents the pion multiplicity $N_{\pi}$ with pion condensation, while the dashed curve represents it without pion condensation. (b) The condensation-spectrum for the VHE gamma ray spectrum, when the highest proton energy $E_p$ does not reach the condensation threshold $E_p^{max}$, the gamma spectrum decays exponentially from $E_{\pi}^{cut}<E_{\pi}^{max}$.
        }\label{fig:3}
	\end{center}
\end{figure}

    We can also measure the gamma-ray spectrum from the condensed $\pi^0$-decay to receive the pion condensation signal.
The gamma-ray spectral energy distribution is $^{[12]}$
\begin{equation}
    \Phi_{\gamma}(E_{\gamma})=C_{\gamma}\left(\frac{E_{\gamma}}{\mathrm{GeV}}\right)^{-\beta_{\gamma}}
\int_{E_{\pi}^{min}}^{\infty}dE_{\pi}
\left(\frac{E_p}{\mathrm{GeV}}\right)^{-\beta_p}N_{\pi}(E_p,E_{\pi})
\frac{d\omega_{\pi-\gamma}(E_{\pi},E_{\gamma})}{dE_{\gamma}},
\label{eq7}
\end{equation}
where the spectral index $\beta_{\gamma}$ is the photon loss due to absorption in the medium near the source. In accelerator
laboratory, the intensity of the proton flux $N_p$ is known, and the power law is taken here for simplicity. $C_{\gamma}$ contains the motion factor and the flux dimension.

    A large number of pions with a certain energy accumulate in a
narrow space during each collision. Due to the overlap of their wave functions, they may transform into each other during the formation time, i.e., $\pi^+ \pi^- \rightleftharpoons 2\pi^0$. However, since $m_{\pi^+} + m_{\pi^-} > 2m_{\pi^0}$ and the lifetime of $\pi^0$ ($10^{-16}$ s) is much shorter than the typical weak decay lifetime of $\pi^{\pm}$ ($10^{-6}$ s - $10^{-8}$ s), the equilibrium will be broken, and $\pi^0$ will dominate the secondary processes, allowing us to neglect the contribution of $\pi^{\pm}$. This result is consistent with the fact: that there is no significant enhancement of the neutrino signal accompanying gamma-ray spectrum, which is a decay product of $\pi^\pm$.  Substituting the standard formula of $\pi^0\rightarrow 2\gamma$ with Equation \ref{eq6} into Equation \ref{eq7}, the following analytical solution is obtained after a simple integration, which is a BPL (Figure 3b)
\begin{equation}
    E_{\gamma}^2\Phi^{GC}_{\gamma}(E_{\gamma})\simeq\left\{
\begin{array}{ll}
\frac{2e^bC_{\gamma}}{2\beta_p-1}(E_{\pi}^{GC})^3\left(\frac{E_{\gamma}}{E_{\pi}^{GC}}\right)^{-\beta_{\gamma}+2} \\ {\rm ~~~~~~~~~~~~~~~~~~~~~~~~if~}E_{\gamma}\leq E_{\pi}^{GC},\\\\
\frac{2e^bC_{\gamma}}{2\beta_p-1}(E_{\pi}^{GC})^3\left(\frac{E_{\gamma}}{E_{\pi}^{GC}}\right)^{-\beta_{\gamma}-2\beta_p+3}
\\ {\rm~~~~~~~~~~~~~~~~~~~~~~~~ if~} E_{\pi}^{GC}<E_{\gamma}<E_{\pi}^{cut},\\\\
\frac{2e^bC_{\gamma}}{2\beta_p-1}(E_{\pi}^{GC})^3\left(\frac{E_{\gamma}}{E_{\pi}^{GC}}\right)^{-\beta_{\gamma}-2\beta_p+3}
\exp\left(-\frac{E_{\gamma}}{E_{\pi}^{cut}}+1\right).
\\ {\rm~~~~~~~~~~~~~~~~~~~~~~~~ if~} E_{\gamma}\geq E_{\pi}^{cut},
\end{array} \right. \label{eq8}
\end{equation}
We call it the condensation-spectrum.

\section{The Observed Data} 

    Unfortunately, all hadron collider experiments in the laboratory have not detected the above condensation signal.
Considering a series of works $^{[13,14,15,16,17,18,19,20]}$ concerning the applications of the condensation-spectrum (Equation \ref{eq8}), we find that the parameters $E_{\pi}^{GC}$ and $E_{\pi}^{cut}$ are target-A dependent in $pA$ collisions. Specifically, $E_{\pi}^{GC}\epsilon [100~GeV, 20~TeV]$ from heavy nuclei to
proton, and it corresponds to $\sqrt{S}\gg 10^6~GeV$ $^{[5]}$. It means that the pion
condensation threshold is beyond the maximum energy available at the accelerator. Therefore, we turn our attention to cosmic rays because their energy may exceed the gluon condensation threshold, where we found that
the condensation spectrum Equation \ref{eq8} can widely interpret very high energy (VHE) gamma-ray spectra in active galactic nuclei, gamma-ray bursts, pulsars, galactic center excess, supernova remnants, even electron/positron and proton/nuclei spectra, where more than seventy astronomical events are recorded $^{[13,14,15,16,17,18,19,20]}$. Now let's add a few more typical examples:
\begin{itemize}
    \item PKS 0625-354$^{[21,22]}$

  This is bright in the X-ray and gamma-ray bands, displaying very high-energy radiation, and is
therefore observed and studied by a wide variety of astronomical equipment. Red-shift $z\sim 0.054$. The variability observed in PKS 0625-354 can be explained by the hadronic model, since changes in the conditions inside the jet (e.g., magnetic field strength, particle density, or acceleration efficiency) lead to rapid changes in the emission. Moreover, the estimation of the radiated power favours the hadronic scenario because of the small proton radiation losses and the higher acceleration efficiency.

    However, the traditional hadronic scenario cannot fit the observed VHE spectrum. Next there was a general preference for
the lepton scenario. The PKS 0625-354 broad-band spectral energy distribution shows two humps that can be interpreted as synchrotron-self-Compton (SSC) emission products. However, a careful review of the lepton programme is not satisfactory. The VHE spectra cannot be described by a single-region SSC scheme. After modification of multiple sources in the low-energy region, different models, the multi-zone SSC scheme and the lepton-hadron SSC scheme can give similar results, suggesting that the correlation between the two peaks is not necessarily inevitable, and that this is a key point in determining whether or not SSC is the radiation mechanism.

  \begin{figure}[H]
	\begin{center}
		\includegraphics[width=1\textwidth]{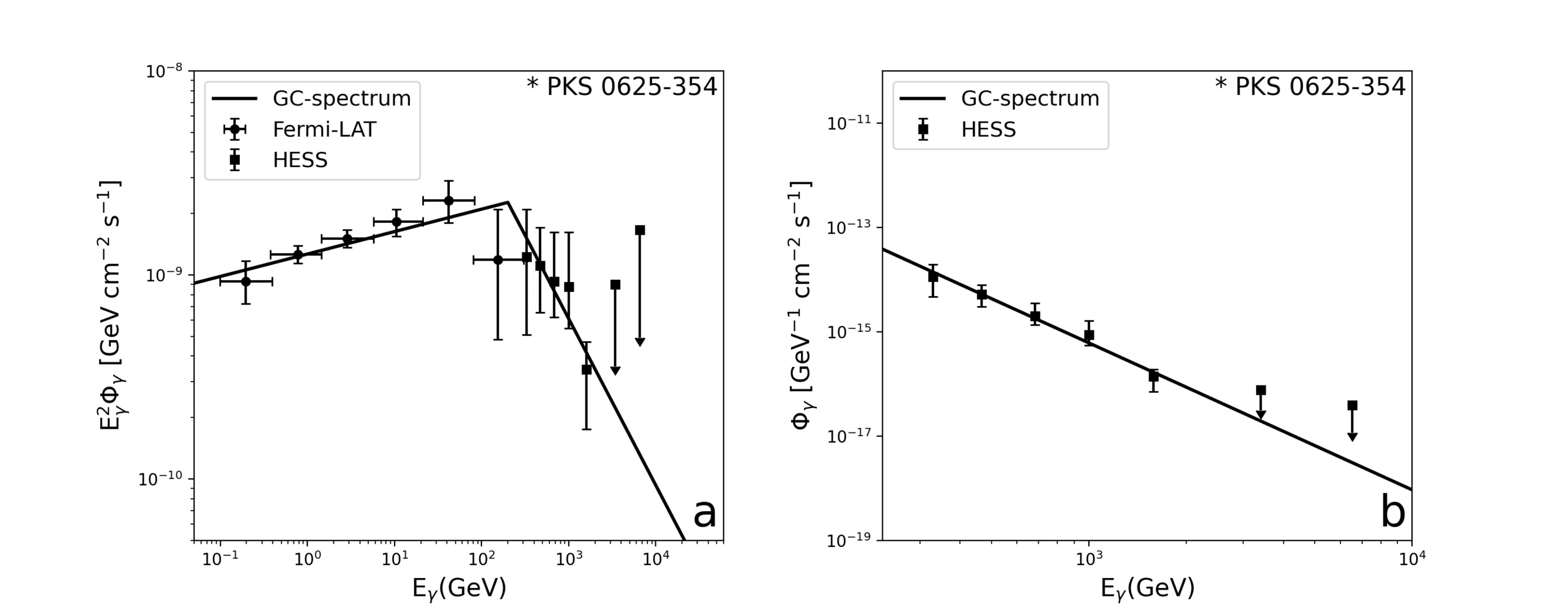} 
		\parbox{15.5cm}{\small{\bf Fig.4.}(a) The condensation-spectrum of PKS 0625-354. Parameters are
$C_{\gamma}=2.57\times 10^{-18}(GeV^{-2}cm^{-2}s^{-1})$, $\beta_{\gamma}=1.89$, $\beta_p=0.98$ and $E_{\pi}^{GC}=200~GeV$.
Data are taken from [21].(b) A part of the condensation-spectrum of PKS 0625-354, which shows the pion condensation signal with $\chi^2/d.o.f.=1.05$.
Data are taken from [22].}\label{fig:4}
	\end{center}
\end{figure}

    Figure 4 is the condensation-spectrum of PKS 0625-354 using Equation \ref{eq8}. It can be found a specific spectrum with
sharp broken power law at $200~GeV$ as shown by the condensation spectrum. The data are corrected for EBL attenuation.
In order to test the PL of $\Phi_{\gamma}$ predicted by pion condensation, Figure 3b was not multiplied by $E_{\gamma}^2$ and
we use Chi-square per degree of freedom ($\chi^2/d.o.f.$) to assess the fitting goodness.

    \item MGRO J1908+06 $^{[23,24]}$
    
    This is an ultra-high-energy gamma-ray source located in the galactic plane. It is notable for
emitting ultrahigh energies in excess of 200 TeV. J1908+06 has a corresponding Fermi-LAT GeV counterpart. The morphology and spectral properties of this counterpart suggest a common origin with the TeV emission, reinforcing the view that the high-energy processes of the source are complex and multifaceted. Figure 5 is an explanation of the condensation-spectrum.

 \begin{figure}[H]
	\begin{center}
		\includegraphics[width=1\textwidth]{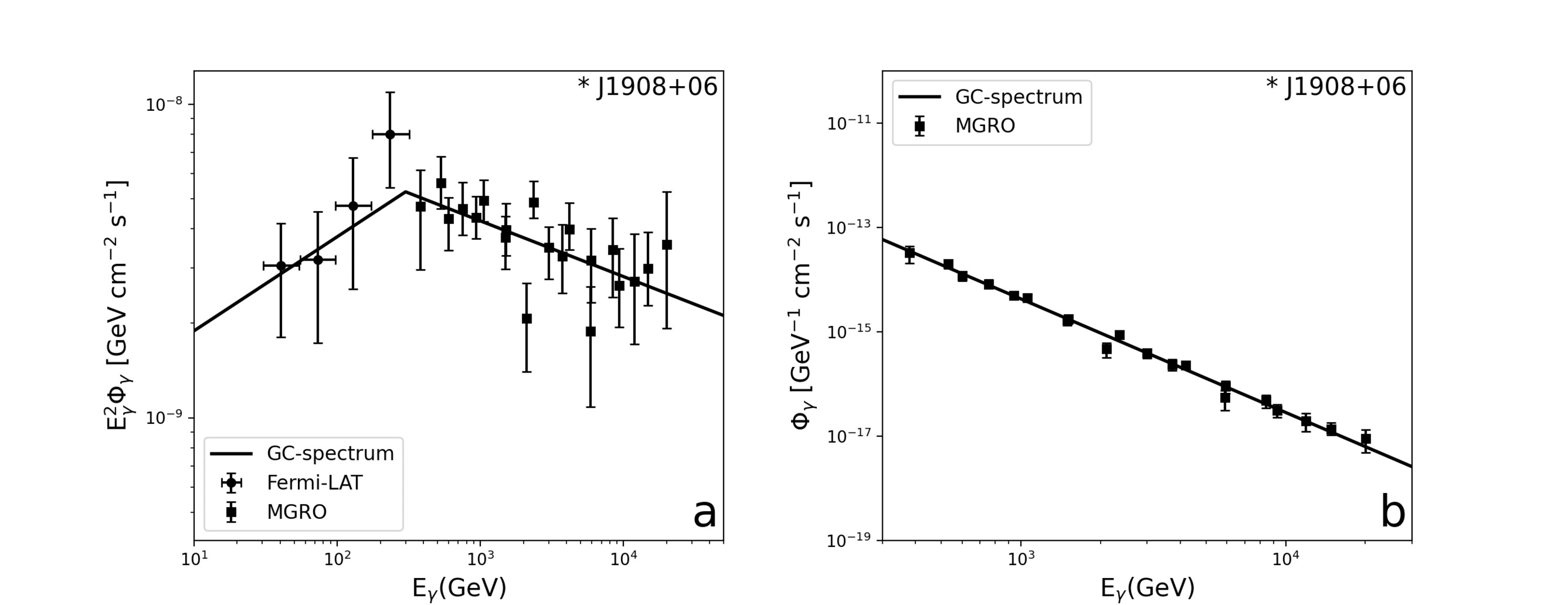} 
		\parbox{15.5cm}{\small{\bf Fig.5.}(a) The condensation-spectrum of J1908+06. Parameters are
$C_{\gamma}=8.9\times 10^{-19}(GeV^{-2}cm^{-2}s^{-1})$, $\beta_{\gamma}=1.70$, $\beta_p=0.74$ and $E_{\pi}^{GC}=300~GeV$.
Data are taken from [23]; (b) A part of the condensation-spectrum of J1908+06, which shows the pion condensation signal with $\chi^2/d.o.f.=0.9$.
        Data are taken from [24].
		}\label{fig:5}
	\end{center}
\end{figure}

\item 1ES 2344+514 $^{[25,26]}$

    BL Lacertae 1ES 2344+514 is a member of the blazar category. It consists of extremely complex physical processes
with extreme variations over time. Since its discovery in 1995, the high-energy peaked blazar 1ES 2344+514 has been observed in the multiwavelength by several other telescopes. Figure 6 shows the GC spectrum of 1ES 2344+514. One can find a clear signal of pion condensation.

\begin{figure}[H]
	\begin{center}
		\includegraphics[width=1\textwidth]{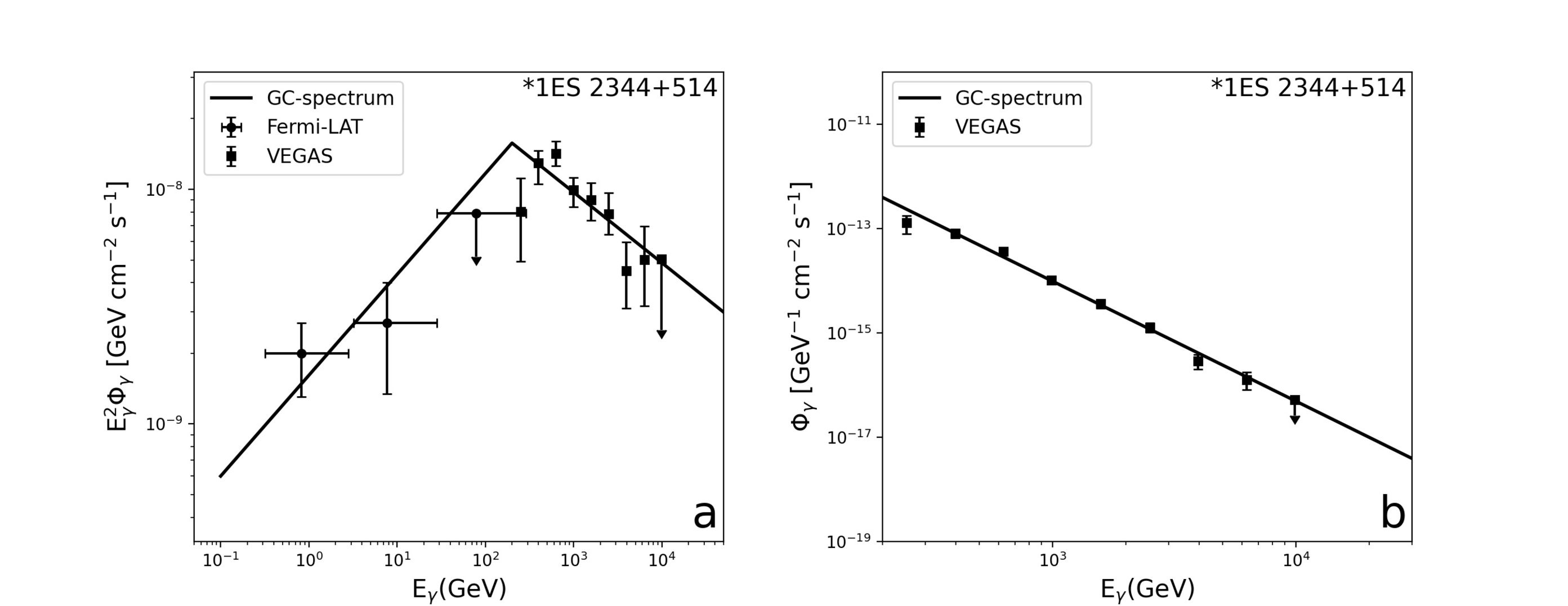}
		\parbox{15.5cm}{\small{\bf Fig.6.}(a) The condensation-spectrum of 1ES 2344+514. Parameters are
$C_{\gamma}=1.38\times 10^{-17}(GeV^{-2}cm^{-2}s^{-1})$, $\beta_{\gamma}=1.57$, $\beta_p=0.87$ and $E_{\pi}^{GC}=200~GeV$.
Data are taken from [25]. (b) A part of the condensation-spectrum of 1ES 2344+514, which shows the pion condensation signal with $\chi^2/d.o.f.=1.4$. Data are taken from [26].
		}\label{fig:6}
	\end{center}
\end{figure}

\item J1356-645 $^{[27,28]}$

    Since the $pp$ collisions are a common phenomenon in the Universe, the pion condensation signals can occur in a variety of environments.
J1356-645 is a known gamma-ray pulsar. Work $^{[28]}$ reanalyze the GeV gamma-ray emission in the direction
of HESS J1356-645 with more than 13 years of Fermi Large Area Telescope (LAT) data. They find that
the spectrum in the energy range in $[1~GeV, 1~ TeV]$ can be described by a PL. It can be explained by the condensation-spectrum (Figure 7).

 \begin{figure}[H]
	\begin{center}
		\includegraphics[width=1\textwidth]{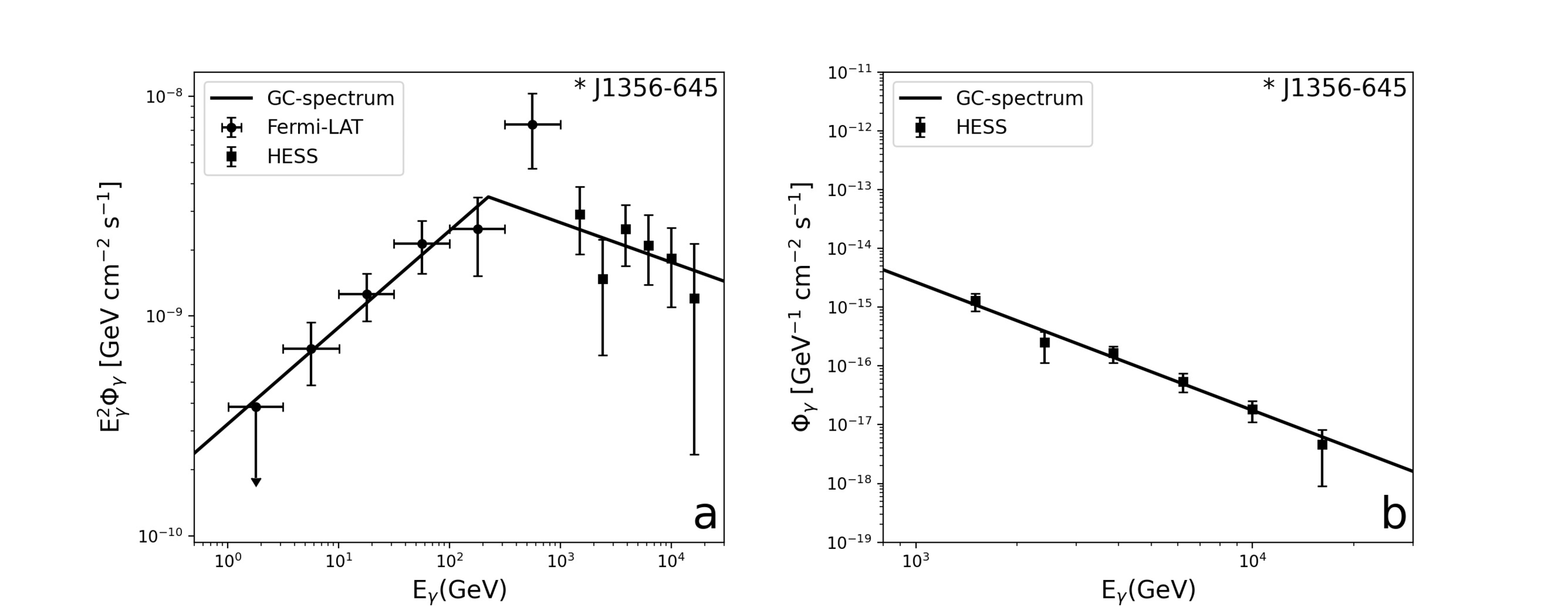}
		\parbox{15.5cm}{\small{\bf Fig.7.}(a) The condensation-spectrum of J1356-645. Parameters are
$C_{\gamma}=1.8\times 10^{-18}(GeV^{-2}cm^{-2}s^{-1})$, $\beta_{\gamma}=1.56$, $\beta_p=0.81$ and $E_{\pi}^{GC}=224~GeV$.
Data are taken from [27]. (b) A part of the condensation-spectrum of J1356-645, which shows the pion condensation signal with $\chi^2/d.o.f.=0.4$. Data are taken from [28].	}\label{fig:7}
	\end{center}
\end{figure}
\end{itemize}

\section{DISCUSSIONS AND CONCLUSION}

     We review the history of BEC. Bose in 1924 proposed a theory about the statistical distribution of
photons, and Einstein found that Bose's method was not only applicable to photons, but also to other particles with integer spins (i.e. bosons). He further proposed that at very low temperatures bosons could condense into the same quantum state to form a completely new state of matter. But it wasn't until 1995 that Cornell and Wieman used laser cooling and magnetic trapping techniques to cool sodium atoms to temperatures close to absolute zero, and succeeded in observing a large number of sodium atoms undergoing the buildup predicted by BEC. The phenomenon of BEC has become an important way to study superfluidity, superconductors, and other quantum phenomena.

     Therefore, cooling dense and unstable pions is a key challenge in realizing pion condensation.
In this work, we point out that the gluons condensed in protons, which was predicted by QCD , can effectively reduce the temperature in the central region of ultra-high energy proton collisions. Thus creating a favorable environment for pion condensation. Although the existing hadron colliders do not have enough energy to produce pion condensation, the prevalence of ultrahigh-energy proton collisions in the Universe is not difficult to create a large number of examples of pion condensation, which can be confirmed by the special shape of the gamma-ray spectrum. We hope this will open a new window for the pion condensation research.

       In summary, we show that once a large number of pions
condense in high-energy hadron collisions, the gamma energy spectra produced by the pion multiplicity and the $\pi^0$-decay show a special PL, the latter of which has already been documented in many astronomical observations, but we have not recognized it. Further analysis supports that the above pion condensation may arise from the gluon condensation in proton. In the $pp$ collisions, a large number of condensed gluons in protons pour into the central region, dramatically increasing the number of pions, which is cooled due to the restriction of energy conservation. In the limit, almost all of the available collision energy is used to create pions, and the central region briefly forms a dense cryogenic region with the potential to form pion condensation. This result not only unravels the mystery of pion condensation, which has been hidden for half a century, but also adds a paradigm for bridging the QCD structure of the proton with cosmic phenomena.

{\bf Acknowledgments:} W.Z. misses Fan Wang dearly for very useful discussions with him. This work is partly supported by
the National Key R$\&$D Program of China (2023YFB3001502) and the National
Natural Science of China (No.12373002).

\end{document}